\documentclass[pdflatex,sn-mathphys]{sn-jnl}

\jyear{2023}%

\theoremstyle{thmstyleone}%
\theoremstyle{thmstyletwo}%

\theoremstyle{thmstylethree}%

\usepackage{csquotes}

\begin{document}

\title[Analyses of prevention trials trajectories.]{How can methods for classifying and clustering trajectories be used for prevention trials? An example in Alzheimer's disease area.}


\author*[1]{\fnm{Céline} \sur{Bougel}}\email{celine.bougel5@laposte.net}

\author[2]{\fnm{S\'ebastien} \sur{D\'ejean}}\email{sebastien.dejean@math.univ-toulouse.fr}

\author[1]{\fnm{Caroline} \sur{Giulioli}}\email{caroline.giulioli.psychologue@gmail.com}

\author[2]{\fnm{Philippe} \sur{Saint-Pierre}}\email{philippe.saint-pierre@math.univ-toulouse.fr}

\author[2]{\fnm{Nicolas} \sur{Savy}}\email{nicolas.savy@math.univ-toulouse.fr}
\equalcont{These authors contributed equally to this work.}

\author[1,3]{\fnm{Sandrine} \sur{Andrieu}}\email{sandrine.andrieu@univ-tlse3.fr}
\equalcont{These authors contributed equally to this work.}

\author[4]{\fnm{the MAPT/DSA group}}

\affil*[1]{\orgdiv{UMR1295 CERPOP - Centre d'Epidémiologie et de Recherche en santé des POPulations}, \orgname{INSERM-University of Toulouse}, \orgaddress{\street{Faculté de Médecine, 37 allées Jules Guesde}, \city{Toulouse}, \postcode{31000}, \country{France}}}

\affil[2]{\orgdiv{UMR 5219}, \orgname{University of Toulouse; CNRS F-31062- UPS IMT}, \orgaddress{\street{Bat 1R3, Université Paul Sabatier, 118 Rte de Narbonne}, \city{Toulouse}, \postcode{31400}, \country{France}}}

\affil[3]{\orgdiv{Clinical Epidemiology and Public Health department, USMR}, \orgname{CHU de Toulouse}, \orgaddress{\street{2 Rue Charles Viguerie}, \city{Toulouse}, \postcode{31300}, \country{France}}}

\affil[4]{Members are listed in the acknowledgments}




\abstract{
\textbf{Background:} Clinical trials are designed to prove the efficacy of an 
intervention by means of model-based approaches involving parametric hypothesis 
testing. Issues arise when no effect is observed in the study population. 
Indeed, an effect may be present in a subgroup and the statistical test cannot 
detect it. To investigate this possibility, we proposed to change the paradigm 
to a data-driven approach. We selected exploratory methods to provide another 
perspective on the data and to identify particular homogeneous subgroups of 
subjects within which an effect might be detected. In the setting of prevention 
trials, the endpoint is a trajectory of repeated measures. In the settings of 
prevention trials, the endpoint is a trajectory of repeated measures, which 
requires the use of methods that can take data autocorrelation into account. 
The primary aim of this work was to explore the applicability of different 
methods for clustering and classifying trajectories. \\
	
\vspace{-0.3cm}	
\textbf{Methods:} The Multidomain Alzheimer Preventive Trial (MAPT) was a 
three-year randomized controlled trial with four parallel arms (NCT00672685). 
The primary outcome was a composite Z-score combining four cognitive tests. The 
data were analyzed by quadratic mixed effects model. This study was 
inconclusive. Exploratory analysis is therefore relevant to investigate the use 
of data-driven methods for trajectory classification. The methods used were 
unsupervised: k-means for longitudinal data, Hierarchical Cluster Analysis 
(HCA), graphic semiology, and supervised analysis with dichotomous 
classification according to responder status. \\
	
\vspace{-0.3cm}		
\textbf{Results:} Using k-means for longitudinal data, three groups were 
obtained and one of these groups showed cognitive decline over the three years 
of follow-up. This method could be applied directly to the primary outcome, the 
composite Z-score with repeated observations over time. With the two others 
unsupervised methods, we were unable to process longitudinal data directly. It 
was therefore necessary to choose an indicator of change in trajectories and to 
consider the rate of change between two measurements. For the HCA method, Ward's 
aggregation was performed. The Euclidean distance and rates of change were 
applied for the graphic semiology method. Lastly, as there were no objective 
criteria to define responder status, we defined our responders based on clinical 
criteria. \\

\vspace{-0.3cm}	
\textbf{Discussion:} In the princeps study, the prevention trial was found to be 
inconclusive, likely due to the heterogeneity of the population, which may have 
masked a treatment effect later identified in a refined subgroup of high Beta 
Amyloid subjects. So, we have adopted an alternative unsupervised approach to 
subject stratification based on their trajectories. We could then identify 
patterns of similar trajectories of cognitive decline and also highlight the 
potential problem of a large heterogeneity of the profiles, maybe due to the 
final endpoint considered.}



\maketitle

\newpage

\section{Background} \label{sec1}

The conclusions of randomized controlled clinical trials are based on clinical and statistical arguments. A trial is considered positive when a significant difference in endpoints is observed between treatment arms and can be interpreted. When this difference is not significant, the trial is deemed inconclusive. This may be due to an ineffective intervention, an insufficiently sensitive method \cite{ROBERTSON2005}, or excessive heterogeneity in the sample. To explore these possibilities, it is useful to study more deeply a particular homogeneous subgroup of subjects that may go unnoticed within this heterogeneous population. To do so, we can characterise subjects who are \enquote{responders} to the intervention, or we can look for homogeneous subgroups of subjects that showed a significant difference between the endpoints observed in different treatment arms. 

Prevention trials, intended to delay the progression or the onset of a disease, have some specific characteristics that make the issue of inconclusive results much more difficult to investigate. First, they often involve longitudinal outcomes \cite{GONG2019}. Second, intervention effects take time to manifest, and are expected to increase with the duration of follow-up. Thirdly, follow-up duration is critical (especially for long-term prevention), as longer follow-ups improve detection but increase dropout risks \cite{TWISK2002attrition, KRISTMAN2004}. The duration of follow-up (a longer duration is preferable, despite the presence of missing values or dropouts), the choice of evaluation criterion (trajectories) and statistical methods are among the key choices in trial design. 
 
For continuous outcomes, trajectory diversity complicates assumptions about curve shapes. Mixed 
effects models or mixed models with latent classes are often used 
\cite{PROUST2013}, though overly supervised by their restricted solution space. Outcome measurement errors further mask the non-linearity of trajectories.
	
Alzheimer’s disease (AD) is particularly suitable for prevention trials 
\cite{WINBLAD2016}, but presents unique design challenges \cite{ANDRIEU2015}. 
The asymptomatic phase of the disease precedes symptoms by several decades (between 10 to 20 years for senile plaques \cite{AMIEVA2009}), affecting follow-up choice. No direct outcome exists for dementia, so surrogate markers like cognitive function must be used \cite{WINBLAD2016}.
	
Most AD prevention trials yield unconvincing results \cite{ANDRIEU2015} if they 
are not studied in more depth through complementary exploratory approaches. It 
is interesting to challenge the classifying and clustering strategies outlined 
below in the context of prevention trials where the endpoints are longitudinal 
and the underlying pattern is non-linear. To identify homogeneous groups of 
subjects, functional unsupervised methods must be used. The more natural way is 
to use unsupervised methods based on distances between trajectories. 
Longitudinal k-means allow this to be done when the number of groups is 
\textit{a priori} specified \cite{GENOLINI2016}. To avoid this issue, a relevant 
alternative is Hierarchical Clustering Analysis (HCA) \cite{JONH1967}. This 
strategy is not specific to functional data, so data pre-processing is necessary 
and is discussed below. To characterize responders (classification), functional 
supervised methods must be used. A responder threshold needs to be defined 
before classification into two groups, which is discussed below. Finally another 
way to classify subjects consists in the use of graphic semiology method 
\cite{HAHS2008}.
	
The aim of this project was to explore clustering and classification tools for longitudinal data, proposing alternative approaches to mixed models that rely on different assumptions to process the data and complement their results. We applied these methods to data from the Multidomain Alzheimer's Disease Prevention Trial (MAPT) and discussed their clinical implications. Finally, we offer recommendations for their use in prevention trials.

\section{Material and Methods}\label{sec2}

All analyses below were carried out using R software (4.1.0, 2021-05-18). The R 
packages used were \texttt{kmlShape} \cite{GENOLINI2016} for longitudinal 
k-means, \texttt{stats} for HCA, and \texttt{seriation} \cite{HAHS2008} and 
\texttt{JLutils} (\url{https://github.com/larmarange/JLutils}) for graphic semiology.
 
    \subsection{MAPT trial}\label{sec2subsec1}
 
The Multidomain Alzheimer Preventive Trial (MAPT) study \cite{ANDRIEU2017} was a 
three-year multicenter, randomized, placebo-controlled superiority trial with 
four parallel arms (allocated by randomization 1:1:1:1). The primary endpoint of 
MAPT was a longitudinal composite Z-score combining cognitive tests, which was 
used to approximate the overall level of cognitive function. The MAPT study was 
analyzed using a quadratic mixed model. The trial concluded that the Multidomain 
Intervention (MI) this means a combination of cognitive stimulation, physical 
activity, nutritional counseling, and preventive consultations \enquote{either 
alone or in combination, had no significant effect on cognitive decline over 
three years in elderly people with memory complaints}. The four arms of the 
study were omega-3 polyunsaturated fatty acids ($\Omega_3$) plus MI, $\Omega_3$ 
alone, placebo plus MI, or placebo alone (control group). Community-dwelling 
subjects aged 70 years or older, without dementia, were treated for 36 months. 
hese subjects had memory complaints, limitations in Instrumental Activities of Daily Living (IADL) \cite{LAWTON1969}, and/or a slow walking speed ($\leq$0.8~m/s), placing them at risk of developing AD. The trial protocol was authorized by the 
Ethical Committee of Toulouse (CPP SOOM III) and the French Health Authority. 
The trial was registered with Clinicaltrials.gov (NCT00672685).
	
This trial postulated that MI with $\Omega_3$ supplementation would have a 
protective effect on cognitive decline, and that the combined intervention would 
have a synergistic effect. To measure global cognitive level, the composite 
score was the average of 4 different Z-scores: the Digit Symbol Substitution 
Test from the Revisited Wechsler code, free and total recall of the Free and 
Cued Selective Reminding test \cite{GROBER1987}, the first 10 items for the 
orientation category of the Mini Mental Status Examination, and the Category 
Naming Test. To obtain it, we summed the Z-scores (standardized by the baseline 
means and standard deviations for each test from our population) of each 
component and divided the total by 4. In \cite{COLEY2016} authors have discussed 
the advantages of considering Z-scores rather than separate scores.
	
The Z-score was assessed at baseline (T0), 6 months (T6), 12 (T12), 24 (T24) and 
36 months (T36) for each of the 1679 subjects~\footnote{Data are available after 
requesting access to the Data Sharing Alzheimer Group – 
info.u1027-dsa@inserm.fr) and are anonymous.}. At each visit, the subjects’ 
cognitive scores for each component of the composite Z-score were recorded 
independently. Subjects were evaluated according to demographic factors (sex, 
age), clinical factors such as symptoms of depression and other cognitive 
performances (Trail Making Test version A (TMT-A) and B (TMT-B) 
\cite{REITAN1958} and the Controlled Word Association Test (COWAT) 
\cite{cardebat1990formal}).
	
    \subsection{Clustering of cognitive function trajectories}\label{sec2subsec2}
 
    \subsubsection{\textbf{k-means for longitudinal data.}}\label{sec2subsec2subsubsec1}
 
Given $k$, a fixed number of groups, k-means is a non-parametric unsupervised 
method for partitioning points into k groups so that the sum of distances 
to the cluster centers (centroids) is minimal. 
Initially, k points are randomly selected as temporary centers. At 
each step, all the other points are assigned to the nearest center, according to 
a proximity criterion, and the cluster centers are updated. This sequential process 
repeats until the partitions remain unchanged, meaning the cluster centers no longer move. 
	
The method was adapted to the longitudinal setting involving a distance 
between trajectories \cite{GENOLINI2011}. k-means for longitudinal data aim to 
define a partition whose trajectories are close at most time points. 
	
With this method, trajectories shifted in time will not be assigned to the same 
cluster. To overcome this issue, the \texttt{kmlShape} package builds clusters 
according to the shape of trajectories regardless of time \cite{GENOLINI2016}. 
The distance between individuals and cluster centers is based on a 
shape-respecting distance (generalized Fr\'{e}chet distance). The centroid 
construction uses a shape-respecting mean (Fr\'{e}chet mean). This specific 
distance, which recalibrates the curves by reparameterising the trajectories, 
tends towards the Euclidean distance when the length of trajectories increases.
	
The second point specific to longitudinal outcomes is missing data. The simplest 
strategy consists in removing subjects with missing values, which results in a 
significant loss of information and a potential selection bias. Another strategy 
is to change the similarity distance with the Gower correction or the 
generalized Fr\'{e}chet distance in \texttt{kmlShape}. Alternatively, missing 
values can be imputed according to usual imputation methods 
\cite{VanBuuren2018}. In \texttt{kmlShape} an imputation method CopyMean is 
available \cite{GENOLINI2013copyMean}. The intermittent missing values were 
calculated using trajectories and an imputation at the time step. Trajectory 
imputation uses linear interpolation, allowing the last available value to be 
linearly linked to the next one. Imputation at the time step uses the average 
imputation to calculate, at the missing point, the average of all values 
available at this point. The methods are then merged to give the final 
imputation value. For monotonic missing values, linear imputation in trajectory 
is replaced by Last Observation Carried Forward, the imputation also uses the 
average of all values available at this point.
	
Furthermore, the \texttt{kmlshape} package requires neither a cluster 
distribution hypothesis nor a shape hypothesis for trajectories. The generalized 
Fr\'{e}chet distance, which considers longitudinal data to be dependent over 
time, does not require the same number of measurements per subject, nor does it 
require measurement times to be identical from one subject to another. However, 
if the number of measurements or trajectories is reduced, then an imputation 
method must be used (linear interpolation by default). Nor does this package 
offer any criteria to select the optimal number of clusters (unlike the 
Calinski-Harabasz criterion \cite{CALINSKI1974} in the \texttt{kml} package).
	
    \subsubsection{\textbf{Hierarchical clustering analysis.}}\label{sec2subsec2subsubsec2}
 
HCA performs the analysis using a set of dissimilarities for the objects being 
clustered \cite{JONH1967}. This is an unsupervised and non-parametric method. 
Each individual is assigned to their own cluster and the algorithm then proceeds 
sequentially to aggregate the two most similar clusters. At each step, the 
similarity is quantified by minimising a prespecified distance (Euclidean, 
Manhattan, Tchebychev, Mahalanobis...). The clustering tree that is finally 
obtained, represented by a dendrogram, provides a hierarchical structure for 
individuals. When the tree is cut toward the leaves (the singletons of the 
individuals), the classification is finer and more homogeneous.
	
The optimal cluster number was chosen \textit{a posteriori} according to the 
inertia criterion. This criterion yields groups with the most similar 
individuals (intra-class homogeneity) and well-defined groups (inter-class 
heterogeneity). To choose the optimal number of clusters, we consider the 
inertia gap of the dendrogram as a function of the number of clusters selected. 
Usually, the partition with the greatest relative loss of inertia is chosen 
(e.g., by using the elbow criterion implemented in the \textit{best.cutree} 
function of the \texttt{JLutils} R library).
	
With longitudinal data, this method cannot be applied directly, as it does not 
take into account the time lag between measurements and assumes independence 
between measurement times. To make the method relevant in our context, an 
indicator of trajectory change has to be defined. Various indicators may be 
considered, such as coefficients of variation of the trajectory, which provide 
a measure of trajectory variation over time for each subject and each 
measurement period, which provide a measure of trajectory variation over time 
for each subject and each measurement period.
 
\subsection{Classification of subjects}\label{sec2subsec3}

Unlike clustering methods, classification methods do not generate groups of 
subjects but rank them in a way that facilitates group formation based on a 
chosen threshold on the outcome. 
	
    \subsubsection{Classification of responders}\label{sec2subsec2subsubsec1}

The key point in classification by respondent status is precisely defining responders. In this supervised approach, a new categorical variable is created based on clinical knowledge (e.g., criteria, group size), with each modality representing a group that can be analyzed or compared. For a continuous variable, multiple thresholds can be used. Standard methods such as description, regression, or machine learning can then be applied to explain these groups. 
 
    \subsubsection{\textbf{Graphic semiology.}}\label{sec2subsec2subsubsec2}

Graphic semiology, or dissimilarities plot, is an alternative to 
HCA. The \texttt{seriation} package uses several seriation and 
sequencing techniques to reorganize datasets or dendrograms, providing cluster representations and visual assessment of cluster tendencies. This combinatorial analysis method, unsupervised and non-parametric, optimizes the reorganization of the data matrix into color shades \cite{HAHS2008}, with rows and columns swapped according to the best permutation. The number of shades has to be chosen before starting and the quantitative values of the dissimilarity matrix are divided into p classes. Each class is given a color shade. A different method can be used 
for rows and columns for more flexibility \cite{HAHS2008}. Up to 30 permutation methods are available, and in the same way as with HCA, an indicator of trajectory change has to be defined.

    \subsection{Samples description}

MAPT trial variables at baseline were used to describe the subgroups of subjects 
resulting from each method. Continuous variables were summarized as mean $\pm$ SD, while categorical variables were reported as counts (percentages). For each variable, comparisons between groups used appropriate statistical tests: Student's t-test for two-group comparisons (with 
normality checked visually), one-way analysis of variance (ANOVA) for means 
comparisons involving more than two groups, and chi-square or Fisher’s exact 
test when validity conditions for proportion comparisons were not perfectly met. 
When ANOVA was used, post hoc comparisons were made using Tukey's honest 
significant difference test. Statistical significance was set at p $<$ 0.05 for all tests.

    \subsubsection{Cognitive tests}
 
Not used in the calculation of Z-scores, two additional cognitive tests were 
used to characterize the groups: the Trail Making Test (TMT) and the Controlled 
Word Association Test (COWAT).
 
The TMT assesses mental flexibility (mainly version B) and information 
processing speed (version A) \cite{reitan1958validity} through a timed task 
involving visual scanning, selective attention, and motor performance 
\cite{COLEY2016}. The score corresponded to its execution time in seconds. 
Performance declines (i.e. execution time increases) with age and is further 
impaired in the elderly or in the presence of Alzheimer's disease 
\cite{bherer2004declin}.

The COWAT measures phonological verbal fluency by assessing the spontaneous 
production of words beginning with a given letter within a 2-minute limit 
\cite{gifford2014source}. Cognitive activity and training have been shown to 
reduce the risk of dementia \cite{COLEY2008}.

    \subsubsection{Clinical factors}

The Clinical Dementia Rating score (CDR) assesses dementia 
status \cite{delrieu2016neuropsychological}, and the degree of severity of cognitive and functional impairment induced by Alzheimer's disease, irrespective of the heterogeneity of the 
pathology. It includes several items such 
as memory, orientation, judgment and problem-solving, activities outside the 
home, domestic and leisure activities and, finally, personal care. In order to simplify the results and make them usable in practice, 
this variable has been considered as binary. For the baseline value, the 
threshold was not important because only 2 classes were present in the MAPT 
population. Moreover, during follow-up, the absence of dementia (score=0) was 
contrasted with the other types of dementia classification (very mild or 
doubtful dementia, mild dementia, moderate dementia, and severe dementia).

The Geriatric Depression Scale (GDS) was a 15-item self-administered 
questionnaire (1 point per item) used to assess depression in 
elderly \cite{sheikh2014geriatric}. It defines three categories: no depression (score $<$ 5 points), probable depression (5–9 points), and highly probable depression ($>$ 9 points). While the link between depression and Alzheimer's disease is widely acknowledged, its nature remains debated \cite{jorm2001history, ownby2006depression}: either as a co-morbidity factor, meaning that affected subjects develop depression 
during the course of their illness; as a precursor to dementia. 
The GDS can be used continuously, with a high score indicating a 
depressive state \cite{AMIEVA2009}. In our study, we dichotomized it at 5 points to distinguish absence vs. presence of depression.

Visual Analog Scales (VAS) assess memory impairment 
experienced by the subject (VAS-1) \cite{COLEY2008} and the discomfort felt in 
daily living activities (VAS-2) \cite{vellas2014mapt}. Subjects mark their perceived severity on a 10 cm scale (left: \enquote{Perfectly} to right: \enquote{Very badly} for VAS-1; left: \enquote{Not at all bothered} to \enquote{Extremely bothered} for VAS-2) by drawing a vertical line. The distance from the left edge is measured in millimeters. Higher scores indicate greater complaints, linked to increased risk of cognitive decline and Alzheimer's disease \cite{vellas2014mapt}. For statistical analyses, VAS-1 and VAS-2 were treated as numerical variables.

Autonomy was assessed using the Alzheimer Disease Cooperative Study - 
Activities of Daily Living Inventory - Prevention Instrument (ADCS-ADL-PI), 
a self-administered questionnaire \cite{vellas2014mapt}. This 15-item scale (0–45 points) measures functional abilities in elderly subjects for Alzheimer's prevention trials. Higher scores indicate greater autonomy, while lower scores reflect loss of autonomy—a key symptom of dementia—used in diagnosis (score decrease) \cite{galasko2006adcs}. This variable was treated as 
numerical.

The Short Physical Performance Battery (SPPB) includes three tasks: walking speed, five chair rises (timed), and balance tests of increasing difficulty \cite{ANDRIEU2017}. Each is scored from 0 (inability) to 4 (best performance), summing to a total score (0–12) that reflects functional status \cite{COLEY2008}. Low SPPB performance is a risk factor for cognitive decline, while regular physical activity may be protective against diseases like Alzheimer’s \cite{COLEY2008}.

Global frailty was assessed using the Fried criterion \cite{fried2001frailty}, which focuses on physical frailty due to reduced physiological reserves. This is part of a broader, multidimensional syndrome encompassing energetic, cognitive, health, and aptitude domains 
\cite{rockwood2005global}. The sum of the 5 criteria leads to a Fried score (0–4 points) \cite{fried2001frailty}, classifying individuals as non-frail (0), pre-frail (1-2), and frail 
(3-4). For our study, a binary variable was created, based on the definition from other studies \cite{buchman2007frailty}, contrasting non-frail (score=0) with other profiles. Each Fried score component is linked to cognitive decline or Alzheimer's disease. Higher scores indicate poorer performance and greater Alzheimer's risk \cite{buchman2007frailty}.

    \subsubsection{Defining criteria used for the post-hoc analysis of MAPT}

The following analyses were conducted after the main study to further explore 
and interpret the results.

Education level reflects socio-cultural and 
quality-of-life factors. Higher education is a well-documented protective factor against dementia \cite{WILS2004}, delaying its clinical 
onset \cite{AMIEVA2009}. In MAPT, it was categorized 
as a binary variable with a 6-year threshold \cite{avila2009cognitive}, distinguishing lower vs. higher education levels \cite{COLEY2008, AMIEVA2009}.

Among the clinical variables, the APOE genotype was considered. Based on 
E2/E3/E4 polymorphisms \cite{liu2013apolipoprotein}, 
subjects were classified into six subgroups (E2/E2, E2/E3, E3/E3, E2/E4, 
E3/E4, and E4/E4) \cite{poirier1993apolipoprotein}.The presence of at least 
one E4 allele (E3/E4 or E4/E4) increases Alzheimer’s disease risk \cite{poirier1993apolipoprotein}. This 
variable thus indicates allele presence/absence.

Cholesterol was collected as a binary variable 
(absence vs. presence) \cite{ANDRIEU2017, liu2013apolipoprotein} and left unchanged. High cholesterol is linked to greater susceptibility to Alzheimer's and cardiovascular diseases \cite{shobab2005cholesterol, liu2013apolipoprotein}. Hypercholesterolemia ($\geq$240 mg/dL) is a known Alzheimer's risk factor \cite{solomon2009midlife}.

Body mass index (BMI) was calculated as weight (kg) divided by height (m²). Higher BMI correlates with increased dementia risk \cite{xu2011midlife}. The rate of dementia increases when BMI increases. 
Typically, an overweight threshold is used, but for consistency across analyses, we used the CAIDE score criterion. Here, obesity was defined as $>$30 kg.m$^{-2}$ \cite{NGANDU2015}.

The Cardiovascular Risk Factors, Aging and Dementia (CAIDE) score estimates Alzheimer's-type dementia risk. It includes seven factors: age ($<$47 years, 47-53 years, and $>$53 years); 
gender (0=female, 1=male); education level (0-6 years, 
7-9 years, and $\geq $10 years); Systolic Blood Pressure 
($\leq$140 mmHg vs. $>$140 mmHg); BMI ($\leq$30 kg.m$^{-2}$, $>$30 
kg.m$^{-2}$); cholesterol levels ($\leq$6.5 mmol/L vs. $>$6.5 mmol/L) 
and Physical Activity (active vs. inactive) 
\cite{kivipelto2006risk}. Since components have different weights, the score ranges from 0 to 15. In population descriptions, the CAIDE score was analyzed both as a composite measure and through its individual components.

\section{Results}\label{sec3}

\subsection{Clustering of cognitive function trajectories}\label{sec3subsec1}

\subsubsection{\textbf{k-means for longitudinal data.}}\label{sec3subsec1subsubsec1}

Since the trajectories started at different times across subjects, we accounted 
for this temporal shift in our analysis. Rather than aligning all trajectories to a common baseline, the \texttt{kmlShape} package clusters individuals based on the shape of their trajectories, rather than their absolute timing. This is achieved 
through the use of the generalized Fréchet distance, which allows for trajectory reparameterization to address temporal misalignment \cite{GENOLINI2016}. The 
\texttt{kmlShape} package was applied to the entire population of 1679 subjects. To ensure reasonable group sizes, we imposed a lower limit of 
10$\%$ of the population for cluster sizes, thereby preventing the formation of excessively small groups. The method was applied directly to the longitudinal composite Z-score values.
	
To initialize the algorithm, two main strategies were considered. The first 
involved applying the \texttt{kmlShape} package directly to the entire population. The 
second involved selecting a subset of the population before applying the method, which could be done either by focusing on specific trajectory shapes or by selecting a subset of individuals for the initial clusters, with the remaining individuals assigned in subsequent steps. Since the results were similar 
regardless of the approach, we chose to apply the method to the entire 
population to avoid introducing an unnecessary selection step. Reducing the 
number of trajectories or individuals can sometimes improve the differentiation of groups or reduce computation time, but in our case, it did not offer 
additional benefits.
	
We retained three groups because the 2-group solution did not provide 
sufficient differentiation between trajectories, as both groups displayed 
relatively stable trajectories without significant clinical differences. 
Conversely, the 4-group solution resulted in one cluster that was too small 
(around 100 subjects), making its interpretation less reliable. 
	
\begin{figure}[h!]
		\centering
		\includegraphics[width=8cm]{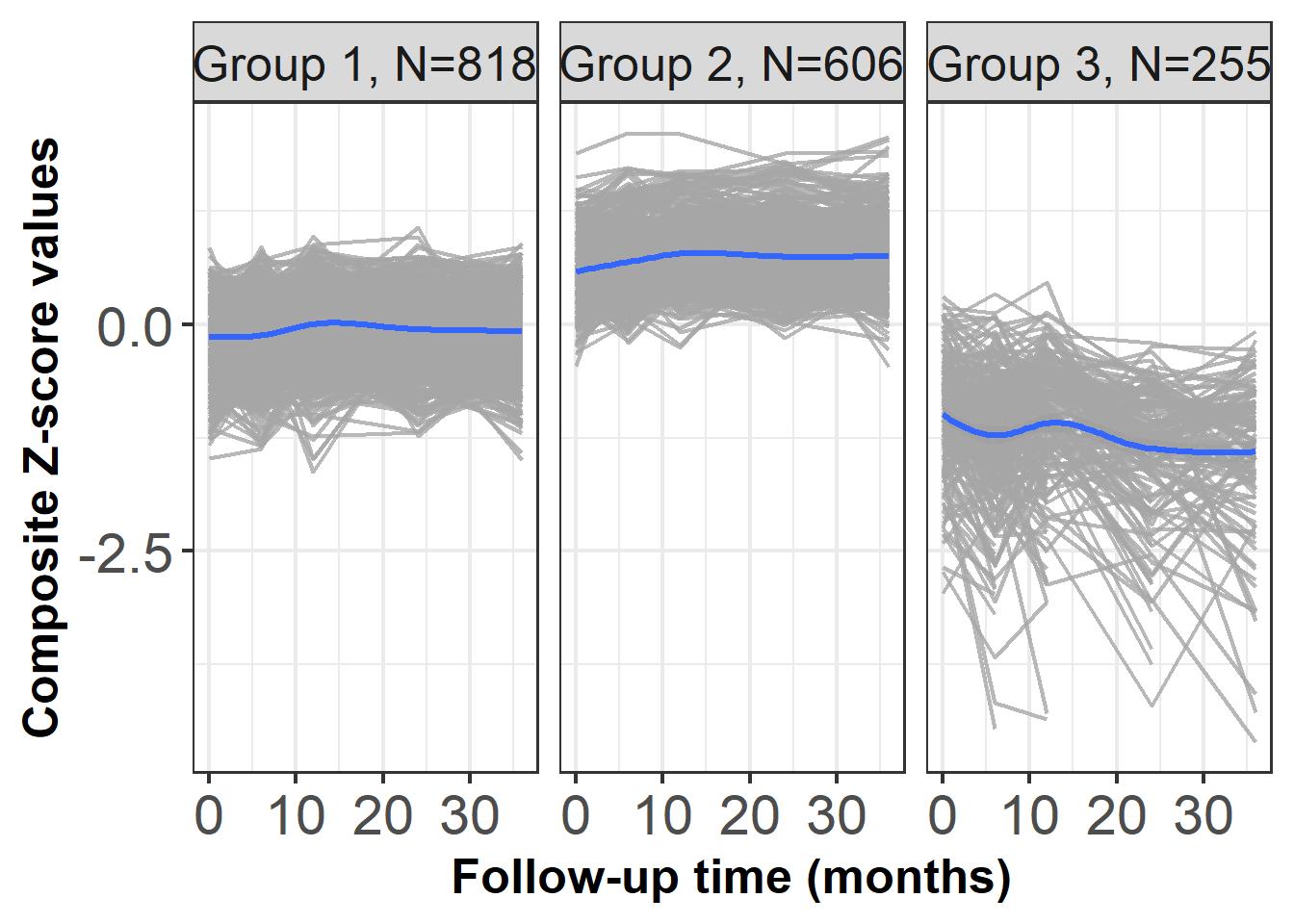}
		\caption{Spaghetti plot of the observed data for the three clusters defined by the kmlShape method for the composite Z-score values, MAPT study (N=1679 subjects). The blue line corresponds to the estimated mean trajectory within each group, based on the composite Z-score values from the MAPT study.}
		\label{kmlShape_image}
	\end{figure}

Following this strategy, the three groups (1, 2 and 3), denoted as G1, G2 and G3, 
included 818 (48.7$\%$), 606 (36.1$\%$), and 255 (15.2$\%$) subjects respectively 
(see Figure~\ref{kmlShape_image}).
	
In examining the clinical characteristics of these groups (data not shown), we 
found that the subjects were equally distributed across the four treatment arms 
of the trial (p=0.30, chi-square test for group distribution). This result was 
consistent with findings from the MAPT trial, where no intervention or treatment effects were observed. However, group G3, which displayed cognitive decline during follow-up, included 64 subjects diagnosed with dementia, representing 92.2$\%$ of all such subjects in the MAPT study. Average 
ages were significantly different between groups, subjects in G3 being older 
(75.5$\pm$4.41 years for G1, 74.0$\pm$3.76 years for G2 and 77.6$\pm$4.83 years 
for G3, p$<$0.001, ANOVA test). Average levels of autonomy also differed significantly between groups, with G3 showing nearly 3 points lower scores compared to the other groups (39.6$\pm$4.55 points for G1, 41.1$\pm$3.83 points for G2, and 
36.6$\pm$6.03 points for G3, p$<$0.001, ANOVA test). Significant differences were observed between groups in memory complaints (average VAS-1 50$\pm$17 mm for 
G1, 47$\pm$16 mm for G2, and 58$\pm$19 mm for G3, p$<$0.001, ANOVA test), 
physical performances (average SPPB at 10.5$\pm$1.6 points for G1, 11.0$\pm$1.4 
points for G2, and 9.9$\pm$2.1 points for G3, p$<$0.001, ANOVA test) and 
cognitive performances (TMT in its both versions and for the COWAT, p$<$0.001 
for all characteristics, ANOVA test). There were also significant differences in the level of education (216 (27$\%$) subjects had a low level of 
education for G1, 58 (10$\%$) for G2, 97 (39$\%$) for G3, p$<$0.001, chi-square 
test). Subjects carrying at least one APOE E4 allele (E3/E4 or E4/E4), associated with an increased risk of Alzheimer's disease, were 
significantly different between groups (143 (23$\%$) for G1, 99 (21$\%$) for G2 
and 57 (31$\%$) for G3, p=0.017, chi-square test). Depressive symptoms also varied significantly across groups (148 (18.2$\%$) subject had positive 
symptomatology according to GDS for G1, 78 (13.0$\%$) for G2 and 72 (28.3$\%$) 
for G3, p$<$0.001, chi-square test). Significant differences were found between groups in the CAIDE score, which was higher in G3 (7.8$\pm$2.0 in G1, 
7.1$\pm$2.0 in G2, 8.12$\pm$1.8 in G3, p$<$0.001, ANOVA test), suggesting a 
greater risk of Alzheimer's-type dementia. This was consistent with the Fried score, which indicates a more frail population (354 (43.3$\%)$ in G1, 221 
(36.5$\%$) in G2, 167 (65.5$\%$) in G3, p$<$0.001, chi-square test), and with the positive CDR (362 (44.3$\%$) in G1, 163 (26.9$\%$) in G2, 181 (71.0$\%$) in G3, 
p$<$0.001, chi-square test). 
 
In summary, subjects in G3 exhibited all the known risk factors associated with cognitive decline.
 
	\subsubsection{\textbf{Hierarchical Clustering Analysis.}}\label{sec3subsec1subsubsec2}
	
We used Ward’s method to aggregate subjects, with Euclidean distance as the 
measure of dissimilarity. This method gives more weight to differences on 
variables with higher variance, as opposed to variables with lower variance. 
In addition, rates of change between each visit were considered as the indicator 
of change. These rates correspond to the difference between two successive values 
of the MAPT composite Z-score, divided by the time elapsed between the two 
corresponding visits. Consequently, the sample had to be restricted to subjects 
with complete follow-up (1143 subjects).
	
\begin{figure}[h!]
		\centering
		\includegraphics[width=8cm]{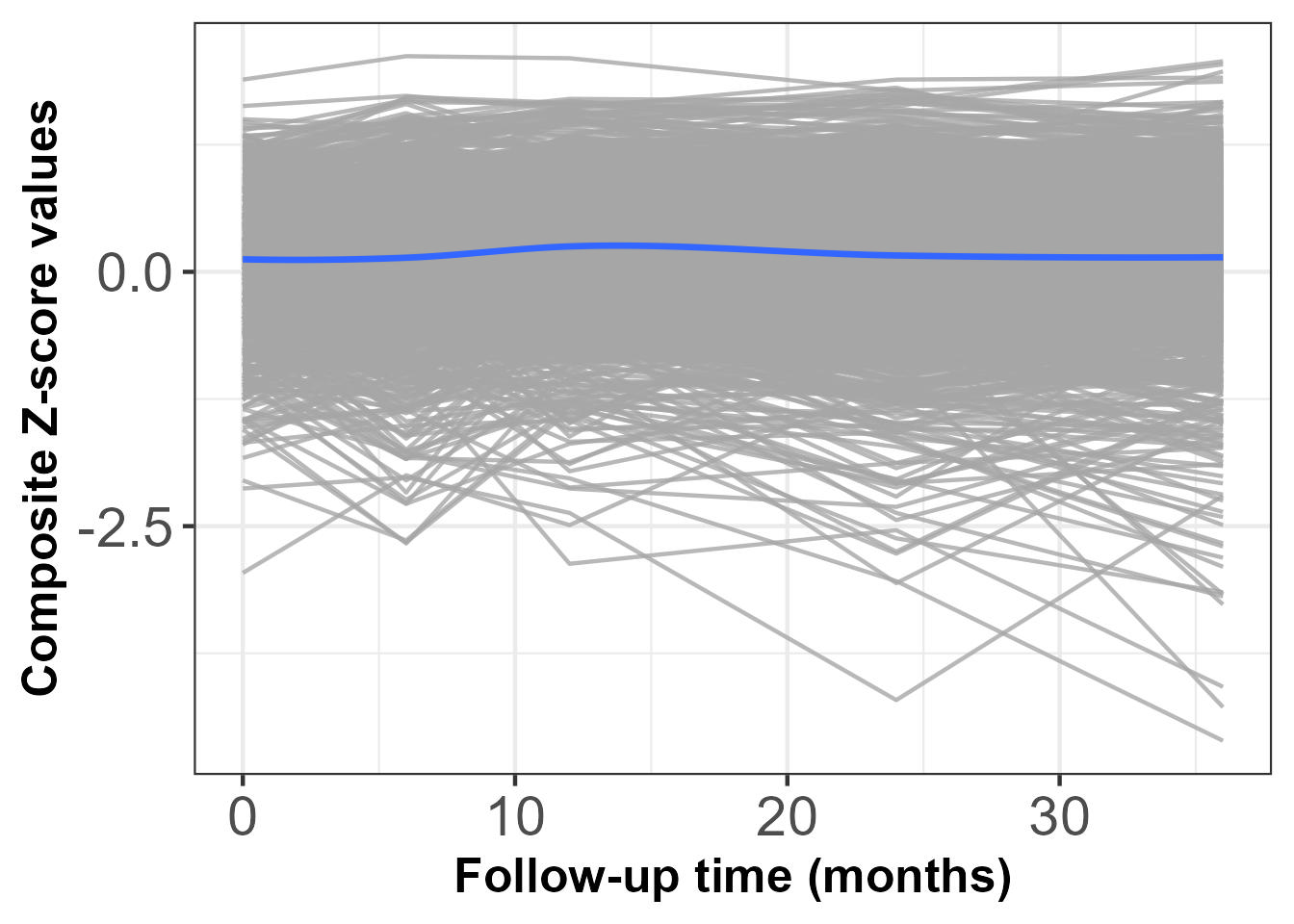}
		\includegraphics[width=8cm]{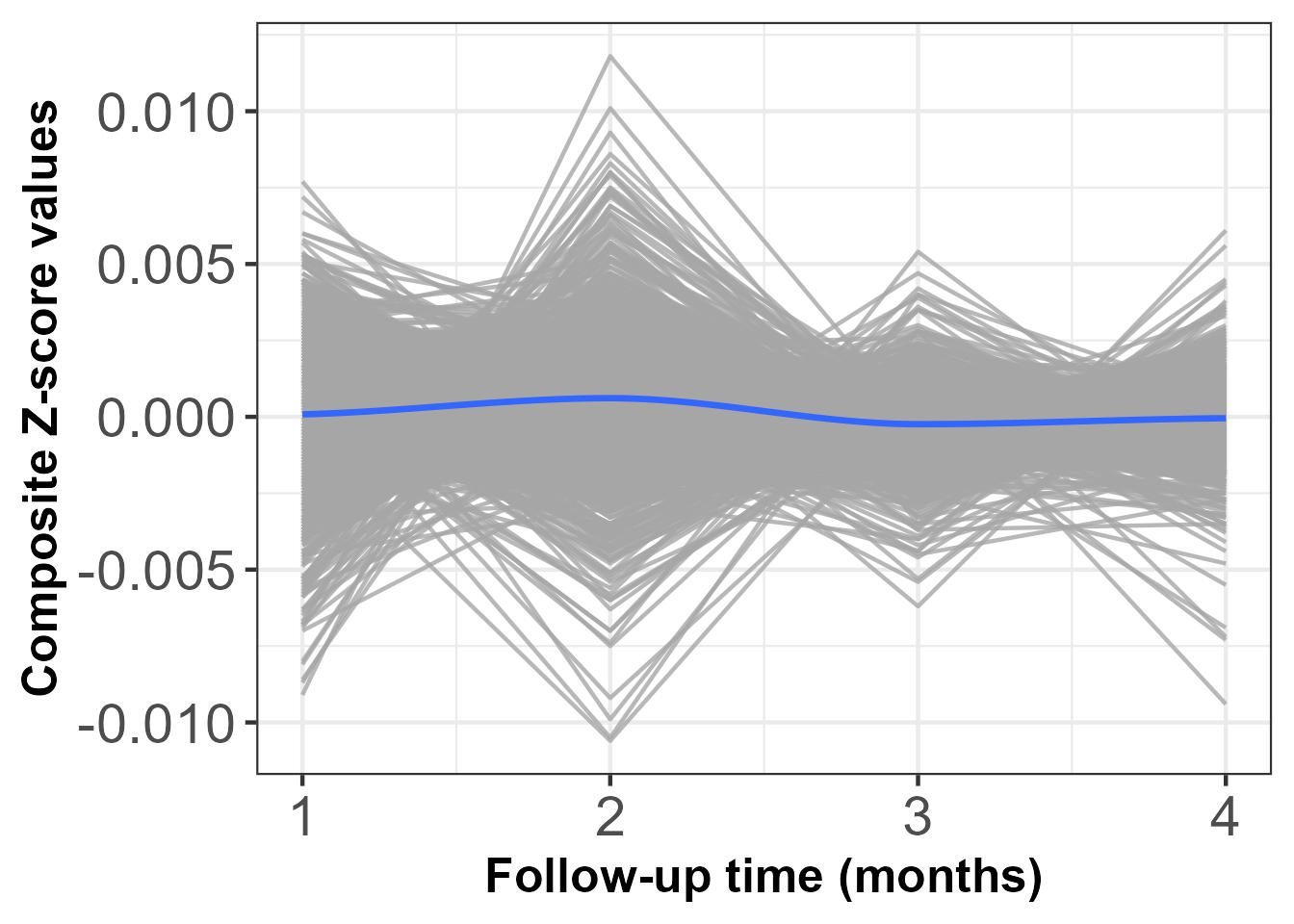}
		\caption{Representation of the composite Z-score values (unprocessed data on the top) and rates of composite Z-score per period (on the bottom) over the follow-up, for the population analyzed, MAPT study (N=1143 subjects). Period 1: T0-T6 months, Period 2: T6-T12 months, Period 3: T12-T24 months and Period 4: T24-T36 months. The blue line corresponds to the estimated mean trajectory within each group. The rates of change were computed separately for each period, ensuring that each individual has four distinct rate estimates. These rates are derived from successive composite score differences divided by the corresponding time interval. Although the values are linked in their computation, the figure displays them as separate estimates for each period, without implying a continuous trajectory between them. The smoothed line represents the overall trend across individuals.}
		\label{MAPT_image}
	\end{figure}

Figure~\ref{MAPT_image} shows differences between changes in the unprocessed 
data and the data defined by rates of change. The average trajectories of the 
unprocessed data were constant during follow-up and were heterogeneous. Looking 
at rates of change over the various periods of follow-up, we quantified the 
change by calculating the rate of increase or decrease between two successive 
measurements. If the rate was non-zero, it indicated a change (non-constant 
data), with data noise appearing to be reduced, except for the T6-T12 month 
period.

For hierarchical cluster analysis (HCA), Euclidean distances were calculated 
based on the rates of change across periods for each individual. These rates 
were treated as separate variables, and their correlations were considered 
during clustering. The hierarchical approach progressively merged individuals 
based on the similarity of their rate profiles, leading to three distinct 
groups that reflect different patterns of change over time.
	
\begin{figure}[h!]
		\centering
		\includegraphics[width=8cm]{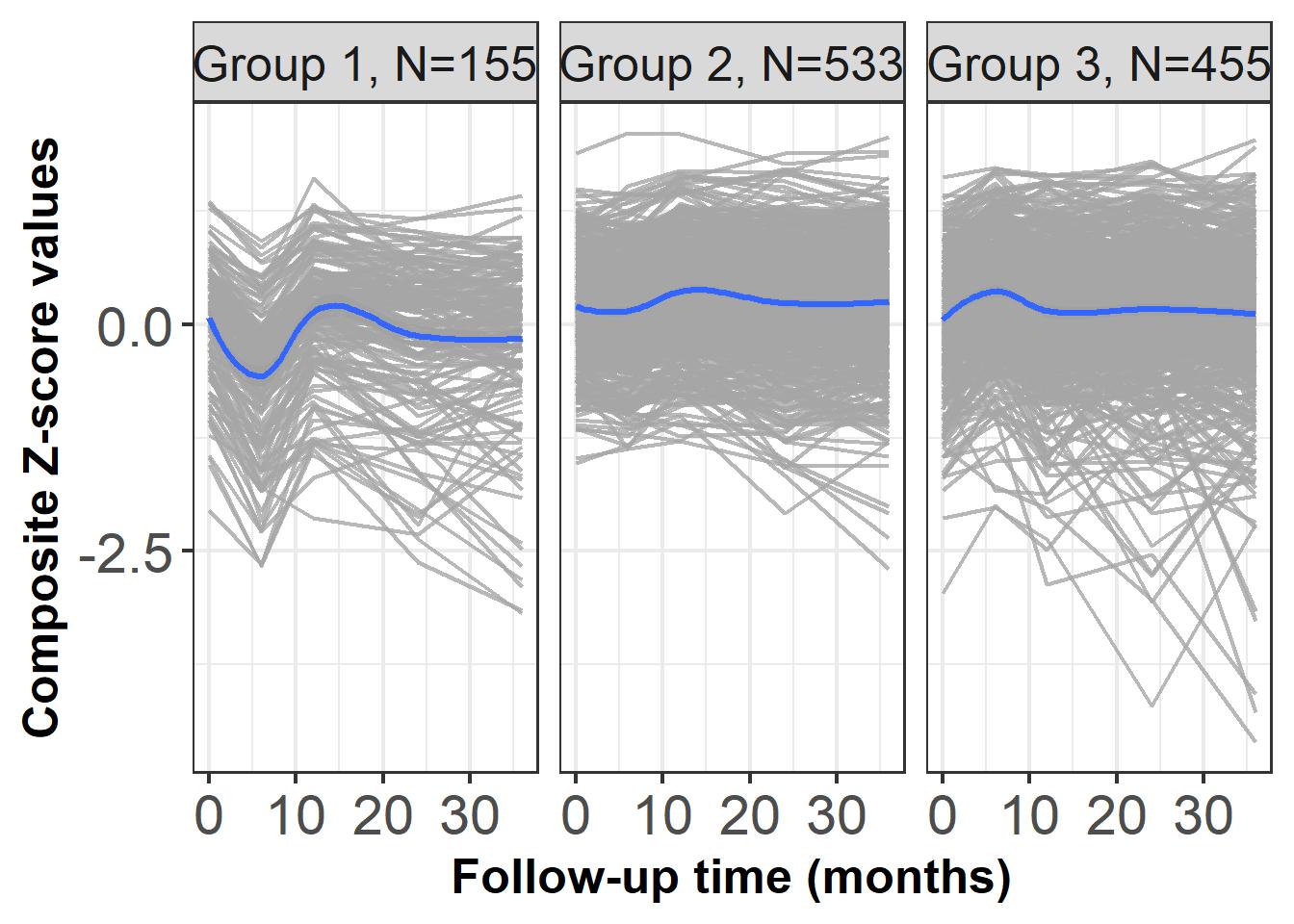}
		\caption{Representation of the different subgroups identified by the Hierarchical Clustering Analysis method, MAPT study (N=1143 subjects). The blue line corresponds to the estimated mean trajectory within each group, based on the rates of composite Z-score values from the MAPT study.}
		\label{HCA_image}
	\end{figure}

Following these choices, HCA led to an optimal partitioning into three groups 
denoted G1, G2 and G3 respectively, as shown in Figure~\ref{HCA_image}. The 
number of subjects in each cluster is shown in the upper part of the figure: G1 (155 (13.6$\%$)) subjects, G2 (533 (46.6$\%$)) 
subjects and G3 (455 (39.8$\%$)) subjects.
 
Clinical characteristics did not differ in distribution between the treatment 
groups (p=0.60, chi-square test for treatment group comparison, data not shown). 
G1 contained fewer obese subjects (12 (7.7$\%$) for G1, 87 (16.4$\%$) for G2 and 
84 (18.5$\%$) for G3, p=0.007, chi-square test) and had a longer TMT-B 
time (122.6$\pm$70.2~s for G1, 113.5$\pm$47.0~s for G2 and 110.7$\pm$46.0~s for 
G3, p=0.006, ANOVA). However, these subjects had fewer difficulties in the 
activities of daily living, as assessed by VAS-2 (34.2$\pm$23.0 mm for G1, 
38.6$\pm$22.3 mm for G2 and 41.2$\pm$24.2 mm for G3, p=0.005, ANOVA, data not 
shown). Finally, subjects in this group had a longer walking time (3.98$\pm$1.1 
in G1, 3.77$\pm$0.9 in G2, 3.81$\pm$0.85 in G3, p=0.036, ANOVA).
	
\subsection{Classification of subjects}\label{sec2subsec3}
	
\subsubsection{Classification of responders}\label{sec3subsec2}

For this analysis, it was essential to clearly define the responder criterion. revious studies have defined responders either based on prior analyses 
\cite{APAYDIN2016, PARASCHAKIS2017} or using traditional definitions such as absolute variation. The most common approach, however, is relative change, which quantifies the difference between baseline and the end of follow-up, divided by the baseline value \cite{GONG2019} or by averaging this difference. 
 	
To perform this analysis, rates of change in the placebo-control group for 
each follow-up period (denoted $Y$) were calculated as the difference between 
the baseline value and the last available value across the full follow-up 
period, excluding missing values during the follow-up. These rates were then compared for each subject (denoted $X$) to those of the placebo-control 
group for the same follow-up period.  If the rate of change for a subject ($X$) was greater than or equal to 20$\%$ compared to the mean rate in the corresponding placebo-control group ($Y$), that subject was considered a responder. The 20$\%$ threshold was chosen based on clinical criteria. A binary variable was subsequently constructed, with a value of 1 for responders and 0 for non-responders. This analysis involved 1145 subjects, excluding those with only baseline data.
	
\begin{figure}[h!]
		\centering
		\includegraphics[width=8cm]{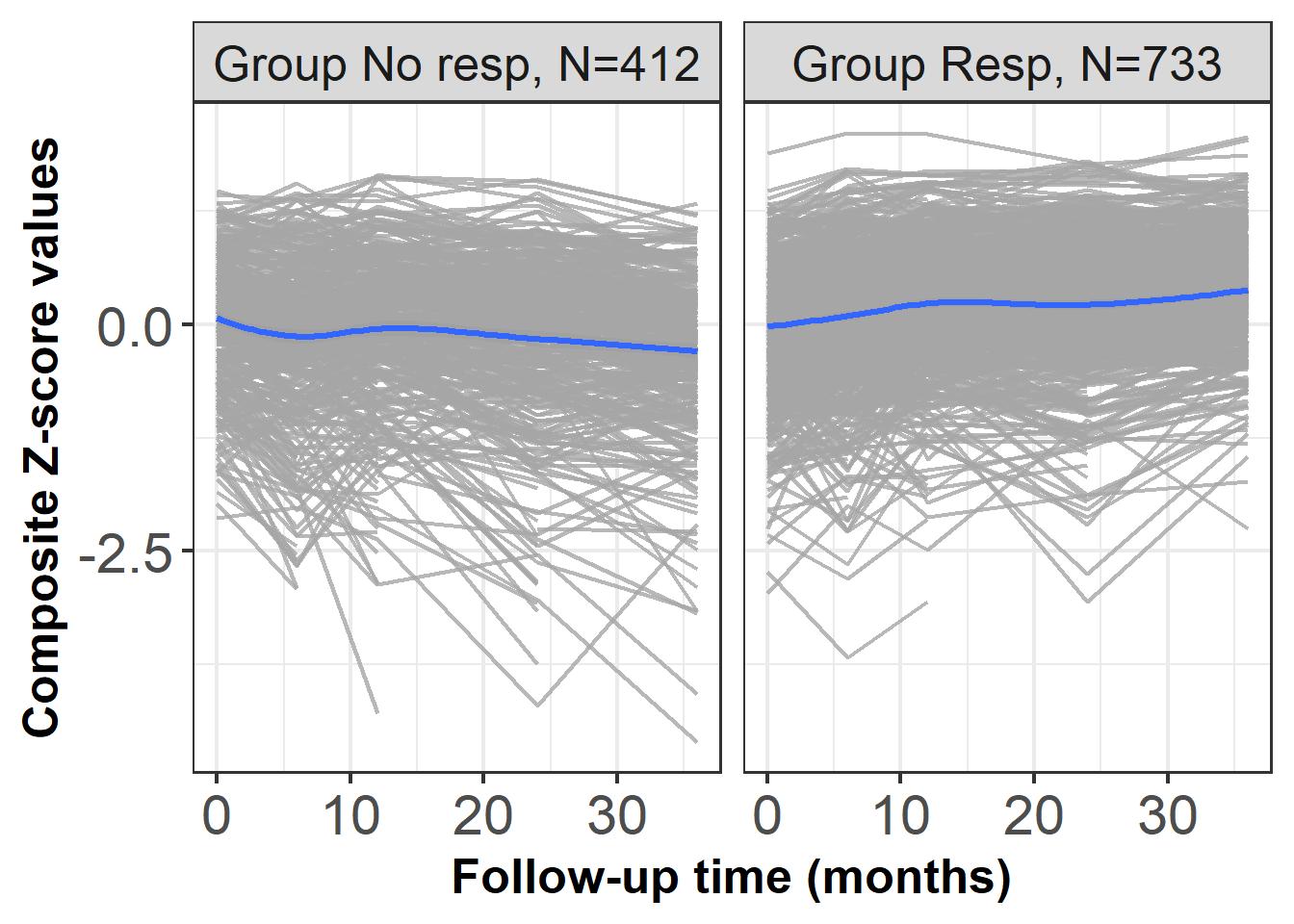}
		\caption{Representation of the different trajectories of the subgroups identified by the responders analysis, MAPT study (N=1145 subjects). The blue line corresponds to the estimated mean trajectory within each group, based on the composite Z-score values from the MAPT study.}
		\label{responders_image_dichotomous}
	\end{figure}

Subjects from the non-placebo groups were classified into two predefined categories. The observed trajectories of these groups are displayed in 
Figure~\ref{responders_image_dichotomous}. The upper portion of the figure shows the number of subjects in each group.
	
Out of the total study population, 733 subjects (48$\%$ of the total or 64$\%$ of the classified subjects) were classified as responders. In other words, nearly a third of subjects improved their cognitive function by 20$\%$ or more compared to the placebo group over the course of three years of follow-up. Characteristics of the groups (data not shown) indicated that responders were younger on average (74.5$\pm$3.87 years vs. 
75.6$\pm$4.37 years for non-responders, p$<$0.001, t-test). Additionally, responders had fewer subsequent diagnoses of dementia during follow-up (17 subjects, 2.3$\%$) compared to non-responders (26 subjects, 6.3$\%$, p$<$0.001, Fisher’s exact test), meaning that 60.5$\%$ of all demented subjects belonged to the non-responder group. Moreover, the responders responder group included a higher proportion of subjects with high cholesterol (47.8$\%$ vs. 38$\%$ for non-responders, p=0.006, chi-square test) and more highly educated subjects (422 (81.2$\%$) vs. 237 
(74.5$\%$), p=0.023, chi-square test). Responders also tended to be less frail (322 (60.9$\%$) vs. 173 (54.2$\%$ ), p=0.057, chi-square test). Finally, responders reported lower perceived discomfort in their daily activities (VAS-2, 37.9$\pm$23.5 mm for responders and 41.1$\pm$22.9 mm for 
non-responders, p=0.046, t-test). However, the distribution of subjects across different treatment arms did not significantly differ between the groups (p=0.20, 
chi-squared test), although the proportion of subjects receiving the multidomain intervention was higher in the responder group (503 (68.6$\%$) in 
responders vs. 261 (63.3$\%$) in non-responders, p=0.069, chi-square test).
 
\subsubsection{\textbf{Graphic semiology.}}\label{sec3subsec1subsubsec3}

We employed a three-color scale: red-white-blue \cite{HAHS2008} to visually represent the 
information carried by the data. The change indicators were rates of change, 
with Euclidean distance applied to the dissimilarity matrix. This method was applied to subjects with complete follow-up data (1143 subjects), as detailed in Section~\ref{sec2subsec2subsubsec2}. o maintain the chronological order of measurements, the columns were kept in the same order using the option \textit{seriate.method=Identity}, meaning that only the rows were rearranged. The default argument \textit{seriate.method=Spectral} was used for row ordering.
	
\begin{figure*}[h!]
		\centering
		\includegraphics[width=11cm]{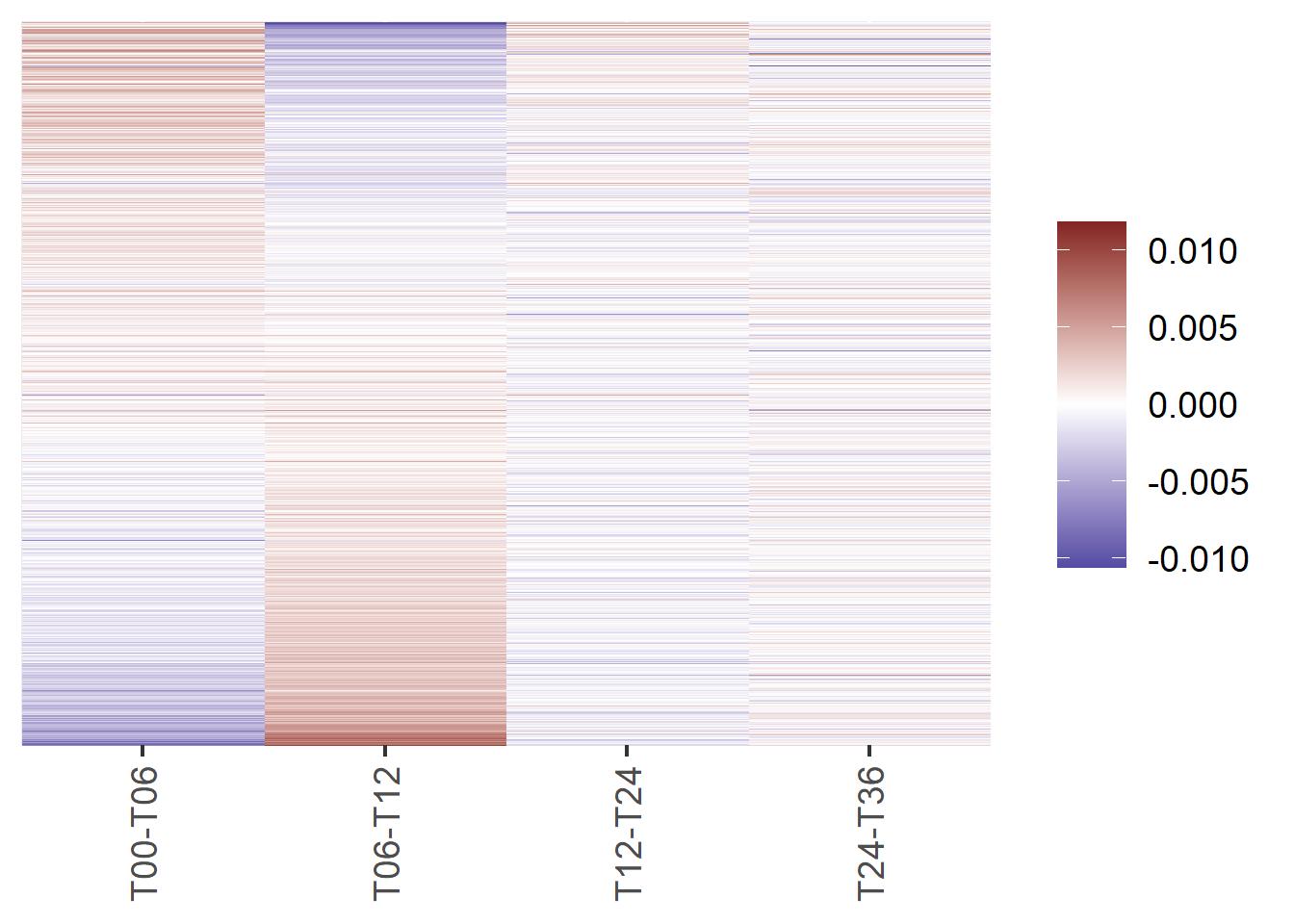}
		\includegraphics[width=8cm]{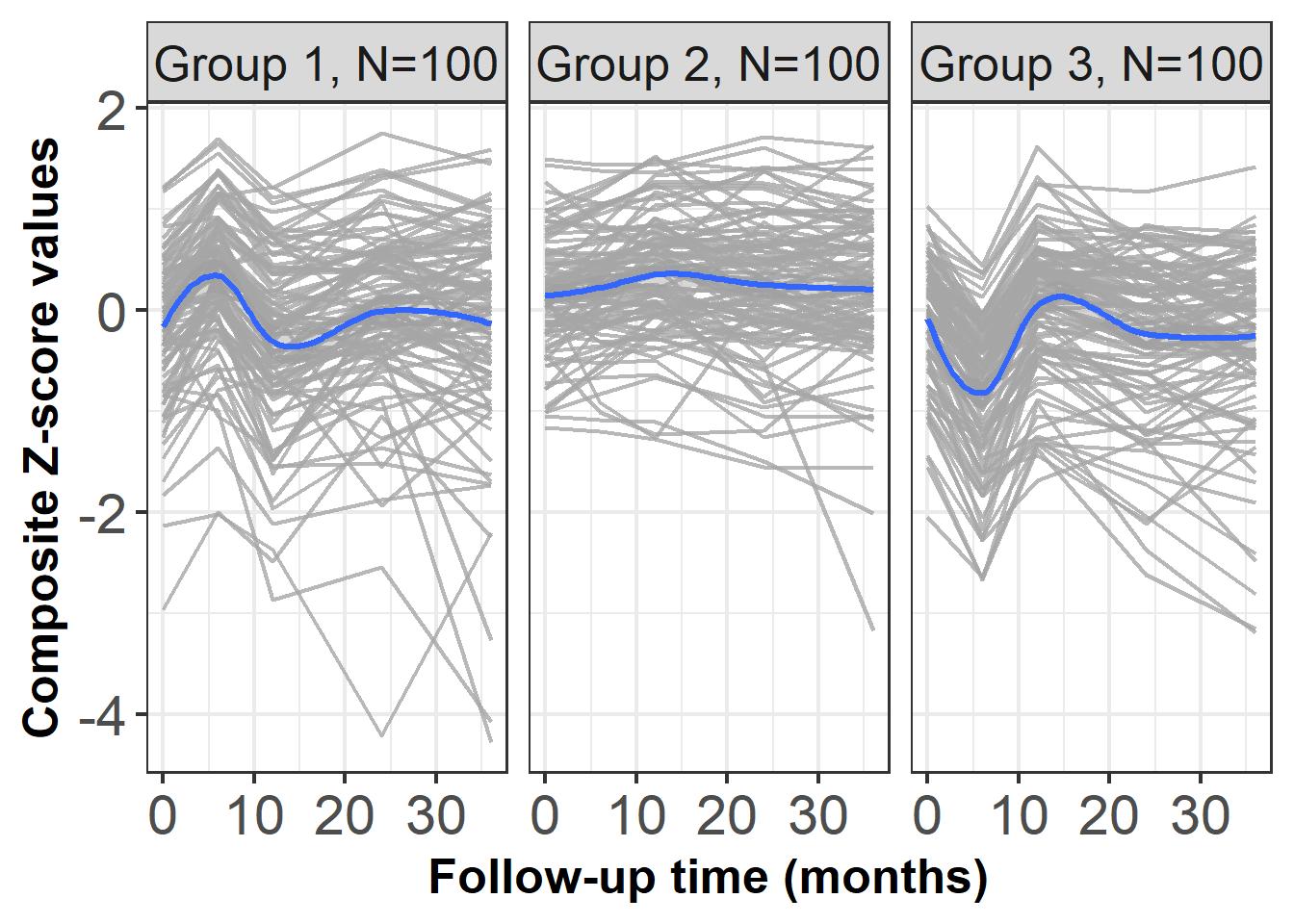}
		\caption{Representation of the color shades of the composite Z-score rates of change identified by seriation (on the top, blue: decrease, white: constant, and red: increase) and spaghetti plot of the observed data for the three subgroups of 100 subjects (on the bottom), MAPT study (N=1679 subjects and N=300 subjects, respectively). The blue line corresponds to the estimated mean trajectory within each group, based on the rates of composite Z-score values from the MAPT study.}
		\label{seriation_image}
	\end{figure*}

Following these choices, permutation analysis revealed a structure 
comprising three groups, as shown in Figure~\ref{seriation_image}, denoted G1, G2 and G3. 
	
Once the database was ordered, groups were manually extracted based on specific thresholds or sample sizes. To illustrate the method, we selected 100 subjects from each group based on their relative positions in the image, measuring the size of the printed representation. Three selection groups were defined: one at the top, one in the middle, and one at the bottom. Using cross-multiplication, we identified the positions of subjects corresponding to these groups. Their trajectories are displayed on the right side of Figure~\ref{seriation_image}. 
	
Clinically (data not shown), there were no significant differences across the treatment arms of the trial (p$>$0.90, chi-square test). However, a difference was noted in baseline TMT-B scores: group G2 (106.3$\pm$40.0s) had the best average score, while group G3 had the worst (140.8$\pm$77.3s), with G1 scoring 124.3$\pm$57.5~s (p$<$0.001, ANOVA). Additionally, more subjects in G1 were classified as obese (21$\%$ of G1, 16$\%$ of G2, and 7$\%$ of G3, p=0.018, chi-square test). A significant difference was also observed in walking speed, with G3 exhibiting the highest speed (4.09$\pm$1.12 m/s), followed by G2 (3.87$\pm$0.86 m/s) and G1 (3.72$\pm$0.84 m/s, p$<$0.05, ANOVA).

\subsection{Overlapping of the groups}

A question of major interest was the overlap between groups identified by the kmlShape, HCA, Seriation, and and LPA (see Supplementary Materials for details on Latent Profile Analysis, LPA) and their relationship with the responder classification. Unlike the previous analysis, we did not restrict the population to a common subset but instead aimed to leverage the maximum available data by constructing contingency tables summarizing individual distributions across methods.

The contingency table between kmlShape and HCA (Table \ref{overlap}) shows relatively low overlap, with the highest correspondence in G2 from kmlShape, representing 40.7$\%$ of G2 in HCA. This suggests that while both methods capture shared structures, they also emphasize different aspects of the data, likely due to dataset heterogeneity, highlighting the complementarity of these techniques.

Our analysis focused on the overlap between responder groups and classifications from kmlShape, HCA, seriation, and LPA. The proportions within each kmlShape group remain stable whether considering the entire population or only individuals with complete composite Z-score data (N=1143), justifying the comparison of responder distributions across methods.

Interestingly, the responder distribution aligns more with HCA-defined groups than with those from kmlShape. Specifically, G3 in kmlShape, representing 15.2$\%$ of the kmlShape population, includes only 5.3$\%$ of responders. In contrast, G3 in HCA, representing 30.1$\%$ of the HCA population, includes 28.8$\%$ of responders. This suggests that the HCA structure may better capture response-related patterns compared to kmlShape.

Further examination shows that G3 from kmlShape does not match any single HCA group but is spread across G3 (44 individuals), G2 (28 individuals), and G1 (34 individuals) in HCA. Conversely, G1 from HCA, which represents 13.6$\%$ of the HCA population, includes 10.0$\%$ of responders and overlaps mainly with G1 from kmlShape (87 individuals). These results indicate that while the clustering methods segment the population differently, HCA-defined groups may be more aligned with response-related characteristics.

The seriation method, which relies on a different clustering principle, also exhibits a distinct responder distribution. Notably, G1 in seriation includes 10.6$\%$ of responders, while G3 includes 7.3$\%$, suggesting that this method identifies a group structure that partially aligns with the responder classification.

The introduction of LPA further refines the comparison of clustering approaches (see Supplementary Materials). While LPA groups exhibit partial overlap with those from kmlShape, HCA, and Seriation, their alignment with responder status provides additional insights. Notably, G1 in LPA shows a higher proportion of responders compared to G3 in kmlShape, reinforcing the hypothesis that LPA captures a distinct structuring of the population. Moreover, the separation between LPA groups suggests a clearer differentiation in cognitive trajectories, further supporting its relevance in this context.

Overall, these findings highlight that different clustering methods segment the population in different ways, reinforcing the importance of using multiple approaches to gain a comprehensive understanding of the data structure. The incorporation of LPA offers an additional perspective on how individuals are classified, particularly regarding their cognitive trajectories and responder status, further supporting the need for a multi-method approach in such analyses.

\section{Discussion}\label{sec12}

In the context of clinical trials, there is no universally recommended method 
for evaluating the results of an inconclusive prevention trial 
\cite{ROBERTSON2005}. While the longitudinal nature of the data, with repeated 
measurements, is often not fully exploited, some trials have used generalized 
mixed models to analyze variations between the beginning and end of follow-up, 
which are well-suited for longitudinal data \cite{ANDRIEU2017, COLEY2016}. 
Other trials have used survival analysis to estimate time-to-event outcomes 
\cite{ROBERTSON2005, COLEY2016}. In the case of the MAPT study, the initial 
hypothesis-driven approach may have been misleading due to the heterogeneity 
of the population, which obscured a treatment effect later identified in a refined subgroup of high Beta Amyloid subjects \cite{DELRIEU2019Reviewer, DELRIEU2023Reviewer}. However, the methods employed and 
the absence of a clear treatment effect were not fundamentally questioned. 

A key point 
of our proposal is that the data could benefit from a more data-driven, a 
posteriori analysis using complementary statistical approaches, such as 
classifying and clustering trajectories. This approach may help better account 
for population heterogeneity and uncover new insights in future analyses.
	
We illustrate the use of four methods that we considered relevant in our context. 
The aim of the study was not to demonstrate the superiority of any individual 
technique over another, but rather to introduce various methods suited for 
longitudinal data, each with its own advantages, with the hope of identifying 
meaningful subject groups. Table~\ref{Summary_methods} summarizes the main 
choices and parameters used for each method, such as model flexibility, data nature, and tuning parameters. This table also includes, for comparison 
purposes, the criteria used in the original study: a mixed model 
estimating the between-group difference in change from baseline to 36 months
\cite{ANDRIEU2017}. The \texttt{kml} package was included in this comparison 
because the \texttt{kmlShape} package, which we used, is derived from it, 
although \texttt{kml} itself was not directly applied. The functional 
clustering \texttt{kmlShape} approach is specifically designed for repeated 
measurements over time, and its distinction from \texttt{kml} lies in its ability to account for trajectory shifts and its approach to determining the number of clusters (using an arbitrary criterion based on group size). 

For the other two unsupervised methods, which required 
preprocessing of the longitudinal data, the key aspect was the choice of the 
variation indicator. This preprocessing step reduces the trajectory to one 
less time point, and we explored different ways to compute it. Some arbitrary choices must also be discussed with clinicians, such as the 
threshold for responders, the relevance of the choice of the variation 
indicator, and the minimal group size, as the results must be clinically 
interpretable.
	
The second objective of this project was to discuss the clinical results 
obtained from the MAPT data. The primary outcome for these 
analyses was a composite Z-score combining four cognitive tests to approximate 
the overall level of cognitive function. Secondary outcomes were the rates 
of change of this score. The \texttt{kmlShape} package highlighted a group where the majority of subjects had dementia, a feature that was also observed, albeit to a lesser extent, in the responder analysis. The proportion of subjects later diagnosed with dementia was lower in the responders group compared to non-responders. Responders also appeared to have clinical 
characteristics more conducive to the absence of cognitive decline.
	
However, subjects diagnosed with dementia could not be 
identified by the other two unsupervised methods, as only complete cases were considered. This population selection, resulting from the definitions of 
the variation indicators, means that different populations were analyzed 
depending on the method used. The results are therefore not directly comparable 
between methods, except in terms of heterogeneity observed across the populations and methods. Importantly, a previous study has 
shown that the more pronounced cognitive decline is associated with earlier loss to follow-up \cite{COLEY2016}.

When we reassess the identification of subjects diagnosed with dementia, we note that, when restricted to the common dataset (N=1143), the proportions of diagnosed subjects across methods vary. Specifically, in the \texttt{kmlShape} approach, the group most associated with dementia diagnoses still included a lower proportion of responders, with G1 containing 66.1$\%$ of responders, compared to the more even distributions observed in the HCA (where G2 contained 67.1$\%$ and G3 contained 63.8$\%$ responders) and Seriation methods, which showed 48.5$\%$ and 54.5$\%$ responders across their groups. This difference may be partly due to the pre-processing steps in the Seriation and HCA methods that excluded incomplete cases, leading to a slightly different pool of subjects. Thus, while these methods provide valuable insights, the difference in populations analyzed—especially with regard to the handling of missing data—limits direct comparisons between them.
	
This work illustrates the utility of multiple methods for clustering and 
classifying subjects. Regardless of the method employed, the aim remains the 
same: constructing meaningful subject groups. The results show that the overlap between these groups is limited, which can be attributed to substantial 
inter- and intra-individual variability, variability over time, and 
the clinical endpoints chosen. Given these findings, 
we suggest that researchers performing post-analysis of inconclusive prevention 
trial data consider using multiple methods and compare the outcomes. To assist 
in this process, Table~\ref{Summary_methods} provides a practical summary of 
the methods and their associated groupings.

\newpage

\section*{Declarations}
\small

\subsection*{Ethical approval}

The trial has been approved by the French Ethical Committee located in Toulouse 
(CPP SOOM II) and was authorized by the French Health Authority.

\subsection*{Informed consent}

All subjects included in the trial were recruited by physicians, who obtained 
written informed consent.

\subsection*{Consent for publication}
All authors of this publication have given their consent for this article.

\subsection*{Declaration of conflicting interests}
	
CB, SD, CG, PSP and NS declare that there is no conflict of interest. SA reports 
a grant and consulting fees from Nestec SA and payment for lectures or 
presentations from Roche.
	
\subsection*{Funding}
This work was supported by the Association Mon\'egasque pour la Recherche sur 
la Maladie d'Alzheimer (AMPA), Harmonie Mutuelle and the Fondation de l'Avenir, 
and by the Alzheimer Prevention in Occitania and Catalonia (APOC) foundation. 
The participation of the Fonds Europ\'een de D\'eveloppement R\'egional (FEDER) 
should also be highlighted. This paper was presented in part at the Journ\'ees 
Ouvertes Biologie, Informatique et Math\'ematiques (JOBIM) in July 2019.
	
\subsection*{Acknowledgements}
	\textbf{MAPT/DSA Group refers to} \\
	MAPT Study Group \\
	\textit{Principal investigator}: Bruno Vellas (Toulouse); \textit{Coordination}: Sophie Guyonnet; \textit{Project leader}: Isabelle Carri\'e; CRA: Laur\'eane Brigitte; \textit{Investigators}: Catherine Faisant, Françoise Lala, Julien Delrieu, H\'el\`ene Villars; \textit{Psychologists}: Emeline Combrouze, Carole Badufle, Audrey Zueras; \textit{Methodology, statistical analysis and data management}: Sandrine Andrieu, Christelle Cantet, Christophe Morin; \textit{Multidomain group}: Gabor Abellan Van Kan, Charlotte Dupuy, Yves Rolland (physical and nutritional components), C\'eline Caillaud, Pierre-Jean Ousset (cognitive component), Fran\c coise Lala (preventive consultation). The cognitive component was designed in collaboration with Sherry Willis from the University of Seattle, and Sylvie Belleville, Brigitte Gilbert and Francine Fontaine from the University of Montreal. \\
	\textit{Co-Investigators in associated centres}: Jean-François Dartigues, Isabelle Marcet, Fleur Delva, Alexandra Foubert, Sandrine Cerda (Bordeaux); Marie-No\"elle-Cuffi, Corinne Costes (Castres); Olivier Rouaud, Patrick Manckoundia, Val\'erie Quipourt, Sophie Marilier, Evelyne Franon (Dijon); Lawrence Bories, Marie-Laure Pader, Marie-France Basset, Bruno Lapoujade, Val\'erie Faure, Michael Li Yung Tong, Christine Malick-Loiseau, Evelyne Cazaban-Campistron (Foix); Françoise Desclaux, Colette Blatge (Lavaur); Thierry Dantoine, C\'ecile Laubarie-Mouret, Isabelle Saulnier, Jean-Pierre Cl\'ement, Marie-Agn\`es Picat, Laurence Bernard-Bourzeix, St\'ephanie Willebois, Il\'eana D\'esormais, No\"elle Cardinaud (Limoges); Marc Bonnefoy, Pierre Livet, Pascale Rebaudet, Claire G\'ed\'eon, Catherine Burdet, Flavien Terracol (Lyon), Alain Pesce, St\'ephanie Roth, Sylvie Chaillou, Sandrine Louchart (Monaco); Kristelle Sudres, Nicolas Lebrun, Nad\`ege Barro-Belaygues (Montauban); Jacques Touchon, Karim Bennys, Audrey Gabelle, Aur\'elia Romano, Lynda Touati, C\'ecilia Marelli, C\'ecile Pays (Montpellier); Philippe Robert, Franck Le Duff, Claire Gervais, S\'ebastien Gonfrier (Nice); Yannick Gasnier Serge Bordes, Dani\`ele Begorre, Christian Carpuat, Khaled Khales, Jean-Fran\c cois Lefebvre, Samira Misbah El Idrissi, Pierre Skolil, Jean-Pierre Salles (Tarbes). \\
	\textit{MRI group}: Carole Dufouil (Bordeaux), St\'ephane Leh\'ericy, Marie Chupin, Jean-François Mangin, Ali Bouhayia (Paris); Mich\`ele Allard (Bordeaux); Fr\'ed\'eric Ricolfi (Dijon); Dominique Dubois (Foix); Marie Paule Bonceour Martel (Limoges); François Cotton (Lyon); Alain Bonaf\'e (Montpellier); St\'ephane Chanalet (Nice); Françoise Hugon (Tarbes); Fabrice Bonneville, Christophe Cognard, François Chollet (Toulouse). \\
	\textit{PET scans group}: Pierre Payoux, Thierry Voisin, Julien Delrieu, Sophie Peiffer, Anne Hitzel, (Toulouse); Mich\`ele Allard (Bordeaux); Michel Zanca (Montpellier); Jacques Monteil (Limoges); Jacques Darcourt (Nice).
	\textit{Medico-economics group}: Laurent Molinier, H\'el\`ene Derumeaux, Nad\`ege Costa (Toulouse). \\
	\textit{Biological sample collection}: Bertrand Perret, Claire Vinel, Sylvie Caspar-Bauguil (Toulouse). \\
	\textit{Safety management}: Pascale Olivier-Abbal \\
	DSA Group \\
	Sandrine Andrieu, Christelle Cantet, Nicola Coley

\newpage

\section*{Tables and figures}

\newpage

\newpage	
	
\newpage
	
\newpage

	\newpage
\begin{table}[h!]
\begin{center}
\begin{minipage}{\textwidth}
\centering
\caption{Repartition of individuals across groups for each clustering method. $N$ represents the total number of subjects considered for each method. The table shows the overlap between groups from different clustering approaches (KmlShape, HCA, Seriation, and LPA), allowing a comparative view of how individuals are classified. The 'Responders' columns indicate the number (proportion) of responders and non-responders within each group, for the shared population across methods. 'NA' stands for 'Not Assigned' or 'Not Applicable' for responders.}
\label{overlap}
\setlength{\tabcolsep}{3pt}
\begin{tabular}{lc|ccc|ccc|ccc|cc|cc}
	\toprule
	\multicolumn{2}{c}{\textbf{Methods}} & \multicolumn{3}{c}{\textbf{KmlShape}} & \multicolumn{3}{c}{\textbf{HCA}} & \multicolumn{3}{c}{\textbf{Seriation}} & \multicolumn{2}{c}{\textbf{Responders}} & \multicolumn{2}{c}{\textbf{LPA}}\\
	& & G1 & G2 & G3 & G1 & G2 & G3 & G1 & G2 & G3 & Oui & Non & G1 & G2 \\
	\cline{2-15}
	\multirow{3}{*}{KmlShape} 
	& G1  & 818 & 0   & 0   & 87 & 262 & 233 & 55 & 50 & 62 & 366 & 189 & 343 & 239\\
	& G2  & 0   & 606 & 0   & 34 & 243 & 178 & 26 & 44 & 12 & 300 & 116 & 0   & 455\\
	& G3  & 0   & 0   & 255 & 34 &  28 &  44 & 19 &  6 & 26 &  67 & 107 & 106 & 0\\
	\midrule
	\multirow{3}{*}{HCA} 
	& G1  &  87 & 34  & 34 & 155 & 0   & 0   & 0   & 0  & 100 &  53 & 65  & 86  & 69\\
	& G2  & 262 & 243 & 28 & 0   & 533 & 0   & 0   & 74 & 0   & 265 & 134 & 177 & 356\\
	& G3  & 233 & 178 & 44 & 0   & 0   & 455 & 100 & 26 & 0   & 211 & 120 & 186 & 269\\
	\midrule
	\multirow{3}{*}{Seriation} 
	& G1  & 55 & 26 & 19 & 0   & 0  & 100 & 100 & 0   & 0   & 49 & 26 & 52 & 48\\
	& G2  & 50 & 44 & 6  & 0   & 74 &  26 & 0   & 100 & 0   & 49 & 25 & 39 & 61\\
	& G3  & 62 & 12 & 26 & 100 & 0  & 0   & 0   & 0   & 100 & 37 & 40 & 64 & 36\\
	\midrule
	\multirow{2}{*}{Responders} 
	& Oui  & 366 & 300 & 67  & 53 & 265 & 211 & 49 & 49 & 37 & 733 & 0   & 178 & 351\\
	& Non  & 189 & 116 & 107 & 65 & 134 & 120 & 26 & 25 & 40 & 0   & 412 & 147 & 172\\
	\midrule
	\multirow{2}{*}{LPA} 
	& G1  & 343 & 0   & 106 & 86 & 177 & 186 & 52 & 39 & 64 & 178 & 147 & 449 & 0\\
	& G2  & 239 & 455 & 0   & 69 & 356 & 269 & 48 & 61 & 36 & 351 & 172 & 0   & 694\\
	\bottomrule
\end{tabular}
\end{minipage}
\end{center}
\end{table}
	
\newpage
\begin{table}[h!]
\begin{center}
\begin{minipage}{\textwidth}
\centering
\caption{Summary, comparisons, and recommendations for using the methods involved in this work. As discussed in Section \ref{sec12}, mixed model and functional clustering using \texttt{kml} methods are not detailed in this paper but for the comparisons to be complete, they have been integrated into this table.}
\label{Summary_methods}
\begin{tabular}{l l l}
    \toprule
    & Mixed model 			&Responder analysis \\
    \midrule
    R Package & \texttt{nlme} 		&None specific package \\
    Data Processed \footnotemark[1] &No 	&RoC between $T_0$ and $T_{\operatorname{end}}$ \\
    Data per subject & 5 		&1 \\
    Usable Dataset &All cases &All except control arm \\
    Imputation\footnotemark[2] & Not necessary 		&Not relevant \\
    Assumption\footnotemark[3] & Shape of trajectory 	&Responder definition \\
    Tuning parameter\footnotemark[4] &Shape fixed by user 			&Threshold fixed by user \\
    \midrule
    & \multicolumn{2}{c}{Functional clustering} \\
    \midrule
    R Package & \texttt{kml} & \texttt{kmlShape} \\
    Data Processed\footnotemark[1] &No &No \\
    Data per subject & 5 & 5 \\
    Usable Dataset & All cases & All cases \\
    Imputation\footnotemark[2] & Not necessary (Grower) & May be relevant \\
    Assumption\footnotemark[3] & Number of groups & Number of groups \\
    Tuning parameter\footnotemark[4] & Calinski-Harabasz & Fixed by group size (10\%)\\
    
    \midrule
    & Graphic semiology & Hierarchical clustering \\
    \midrule
    R Package & \texttt{seriation} & \texttt{stats} \\
    Data Processed\footnotemark[1] &RoC between visits &RoC between visits \\
    Data per subject & 4 & 4 \\
    Usable Dataset & Complete cases & Complete cases \\
    Imputation\footnotemark[2] &Not relevant &Not relevant \\
    Assumption\footnotemark[3] &Number of shades & Number of groups \\
    Tuning parameter\footnotemark[4] &Fixed by user &Elbow criterion \\
    \botrule
    \end{tabular}
\end{minipage}
\end{center}
\footnotetext[1]{The outcome of interest is MAPT composite Z-score. Depending on the method, the data used can be whole the trajectories or data processed by means of indicator of changes. In this table one considers the rate of change (RoC) but alternative indicators may be used: min, max, cumulative variation, sum of absolute variations...}
\footnotetext[2]{To deal with missing data, it may be of interest to use data imputation methods. However, some methods are able to handle missing data, and thus imputation is not necessary. Furthermore for some methods data imputation is not relevant.}
\footnotetext[3]{Depending on the methods there are underlying assumptions made. These assumptions concretizes by parameter(s) to be fixed by the user before analyses.}
\footnotetext[4]{Specification of the method for choosing the parameter(s) of the underlying assumptions.}

\end{table}

\end{document}